\input epsf.tex
\def\DESepsf(#1 width #2){\epsfxsize=#2 \epsfbox{#1}}
\documentclass{ws-procs9x6}

\begin{document}

\title{Quark, Lepton and Neutrino Masses In Horava-Witten Inspired Models}

\author{R. Arnowitt, B. Dutta\footnote{Present address: Department of Physics, University of Regina, Regina
SK, S4S  0A2 Canada} and B. Hu}

\address{Center For Theoretical Physics, Department of Physics, Texas A$\&$M University,
College Station TX 77843-4242, USA\\ 
E-mail: arnowitt@physics.tamu.edu, duttabh@uregina.ca, b-hu@physics.tamu.edu}


\maketitle

\abstracts{ Horava-Witten M-Theory offers new ways in achieving the quark and lepton mass
hierarchies not naturally available in supergravity unified models. In
previous work, based on a torus fibered Calabi-Yau manifold with
a del Pezzo base dP$_7$,a three generation SU(5)model with Wilson line breaking
to the Standard Model was constructed. It was seen that if the additional 5-
branes clustered near the distant orbifold plane, it was possible that such
models could generate the observed hierarchies of quark masses without undue
fine tuning. We update these results here and extend them to include the
charged leptons. A new mechanism for generating small neutrino masses
(different from the ususal seesaw mechanism)arises naturally from a possible
cubic holomorphic term in the Kahler potential when supersymmetry is broken. 
We show that the LMA solution to neutrino masses can occur, with a good fit to all neutrino
oscillation data. The model implies the existance of the 
$\mu \rightarrow e \gamma$ decay even for
universal slepton soft breaking masses, at rates accessible to the next round
of experiments.}

\section{Introduction} Over the past year considerable progress has been made in
understanding Horava-Witten heterotic M-theory \cite{hw1} with ``non-standard"
embeddings\cite{o}. In this picture, space has an 11
dimensional orbifold structure of the form (to lowest order) $M_4\times
X\times S^1/Z_2$ where $M_4$ is Minkowski space, $X$ is a 6D
Calabi-Yau space, and $-\pi\rho\leq x^{11}\leq\pi \rho$. The space thus has two
orbifold 10D manifolds $M_4\times X$ at the $Z_2$ fixed points at $x^{11}=0$
and $x^{11}=\pi\rho$ where the first is the visible sector and the second is
the hidden sector, each with an apriori $E_8$ gauge symmetry. In addition
there can be a set of 5-branes in the bulk at points
$0< x_n<\pi\rho$, $n=1...N$ each spanning $M_4$ (to preserve Lorentz
invariance) and wrapped on holomorphic curve in $X$ (to preserve N=1
supresymmetry).

In general, physical matter lives on the $x^{11}=0$ orbifold plane and only
gravity lives in the bulk. The existence of 5-branes allows one to satisfy the
cohomological constraints with $E_8$ on $x^{11}=0$ breaking to $G\times H$
where $G$ is the structure group of the Calabi-Yau manifold and $H$ is the
physical grand unification group.  We consider the case
$G=SU(5)$, and $H=SU(5)$.

In this picture, three generation
models with Wilson line breaking $SU(5)$ to $SU(3)\times SU(2)\times U(1)_Y$
have been constructed using torus fibered Calabi-Yau
manifolds (with two sections)\cite{dopw}. Also, the general structure (to the
first order) of the Kahler metric of the matter field has been
constructed\cite{low4}. In general such models do not naturally give rise to
observed quark/lepton mass hierarchies. It was shown\cite{ad}, howevere, that
models with vanishing first pontrjagin class on the physical orbifold,
$\beta_i^{(0)}$ can lead to Yukawa textures with all CKM and quark mass data
without any undue fine tuning, and three generation model with wilson line
breaking of $SU(5)$ to SM exists provided the Clabi-Yau manifold has del Pezzo
base $dP_7$. ( The possibility that vanishes is non-trivial since it is
forbidden in elliptically fibered Calabi-Yau manifold\cite{ow}.)
Within this framework, in this note, we evaluate the charged lepton masses, neutrino masses and the
mixing angles and explore the signal of this model in the rare decay modes of leptons.

\section{Kahler Metric} The bose part of the 11 dimensional gravity multiplet
consists of the metric tensor $g_{IJ}$, the antisymmetric 3-form $C_{IJK}$ and
its field strength $G_{IJKL}$ ($I,\,J,\,J,\,K=1...11.$).
The $G_{IJKL}$ obey field equations $D_IG^{IJKL}=0$ and Bianchi identities
\begin{eqnarray}\label{eq3}(dG)_{11RSTU}&=&4\sqrt 2\pi({\kappa\over
{4\pi}})^{2/3}[J^0\delta(x^{11})+J^{N+1}\delta(x^{11}-\pi \rho)\\\nonumber &+&{1\over
2}\Sigma^N_{n=1}J^{n}(\delta(x^{11}-x_n)+\delta(x^{11}+x_n))]_{RSTU}
\nonumber\end{eqnarray}
Here $(\kappa^{2/9})$ is the 11D Planck scale($k_{11}$), and $J^{n}$,
$n=0,\,1,\,...N+1$ are sources from orbifold planes and the $N$ 5-branes.
These equations can be solved perturbatively in powers of $(\kappa^{2/3})$
\cite{low4}. The effective 4D theory can then be characterized by a Kahler
potential $K=Z_{IJ}{\bar C^I}C^J$, Yukawa couplings $Y_{IJK}$ for the matter
fields $C^{I}$ and gauge functions from the physical orbifold plane $x^{11}=0$.
To first order, $Z_{IJ}$ takes the form \cite{low4}. 
\begin{equation}  Z_{IJ}=e^{-K_T/3}[G_{IJ}-{\epsilon\over{2V}}\tilde{\Gamma}^i_{IJ}
\Sigma^{N+1}_0(1-z_n)^2\beta^{(n)}_i]
\label{eq2}\end{equation}
\section{Phenomenological Yukawa Matrices} One expects $G_{IJ}$, $\tilde\Gamma^{i}_{IJ}$ and
$Y_{IJK}$ to be characteristically of $O(1)$, and the parameter $\epsilon$ is
not too small. However, the second term will be small if $\beta^{(0)}_i$ were
to vanish and if the 5-branes were to be near the distant orbifold plane i.e.
$d_n\equiv1-z_n$ were small. In the following we will assume then that
$\beta^{(0)}_i=0;\,\,d_n=1-z_n\cong0.1$.

 A phenomenological
example for the quark contributions with these properties (and containg
the maximum numbers of zeros) is $(f_T\equiv exp(-k_T/3))$:
\begin{equation} Z^u=f_T\left(\matrix{ 1  & 0.345  & 0 \cr 0.345 & 0.131 & 0.636 d^2
\cr 0  & 0.636 d^2& 0.34 d^2 }\right); \,\,\,Z^d= 
f_T\left(\matrix{ 1 & 
0.49  & 0 \cr 0.49 & 0.56 & 0.43 \cr 0  & 0.43& 0.72  }\right).
\label{eq27}\end{equation}
with Yukawa matrices ${\rm diag}Y^u={(0.01,\, 0.06,\, 0.1 Exp[
0.65 \pi i ])}$ and ${\rm diag}Y^d={(2,\, 0.25,\, 1.82)}$. 

Thus to obtain the physical Yukawa matrices, one must first
diagonalize the Kahler metric and then rescale it to unity. Then using the
renormalization group equations, one can generate the CKM matrix, and the quark
masses. We use $\tan\beta=40$ in our analysis. We obtain the correct quark
masses and mixing angles. We also get the CP violating parameter $sin2\beta=0.8$
which agrees well with the experimental data \cite{Babar}.

In order to obtain correct charged lepton masses, we need:\begin{equation} 
Z^l=f_T\left(\matrix{ 1  & -0.547  & 0 \cr -0.547 & 0.43 & 0.025
\cr 0  & 0.025& 0.09 }\right); Z^{e_R}=f_T\left(\matrix{ 1  & 0.624  & 0 \cr
0.624 & 0.4 & 0.57d^2
\cr 0  & 0.57d^2& 0.44d^2 }\right) \end{equation}
and  Yukawa matrix ${\rm diag}Y^l={(0.3,\, 3,\, 1.82)}$.

\section{neutrino masses and mixing angles}The small neutrino masses and the
near bi-maximal mixing (as required by the recent experimental data\cite{neutrino}) arise from a Kahler
potential term in this model:
$k_{11} /{f_T^{3/2}\sqrt{G_{H_2}}}\lambda_{\nu} L\nu_RH_u
$. Under a Kahler transformation, we get the following term in the superpotential:\begin{equation}
{k_4^2<W_h>k_{11}\over{f_T^{3/2}\sqrt{G_{H_2}}}}\lambda_{\nu} L\nu_RH_u, \end{equation}where $k_4^2=8\pi G_N$. $W_h$ is the hidden sector superpotential.
Now using  \begin{equation}
Z^{\nu}=f_T\left(\matrix{ 1  & -0.465  & 0 \cr -0.465 & 0.31 & 0.025
\cr 0  & 0.025& 0.028 }\right) \end{equation}
with ${\rm diag}Y^l={(4,\, 0.1,\, 4)}$,  we generate the following
neutrino masses: $(2.4\times 10^{-3},\,1.4\times 10^{-2}, \,1)x$, where
$x=k_4^2<W_h>k_{11}/{f_T^{3/2}\sqrt{G_{H_2}}}$ 
and mixing angles: $
\tan^2\theta_{12}=0.41,\, \tan^2\theta_{23}=0.89.$ 
\section{Lepton flavor violation}Since the physical lepton couplings are flavor
non diagonal, we get appreciable lepton flavor violating decay  of $\mu\rightarrow
e\gamma$. We use the minimal SUGRA framework i.e. universal scalar, gaugino
masses and trilinear A terms at the GUT scale to calculate the branching ratio
(BR) of
this decay. We plot the BR as a function of $m_{1/2}$ in Fig.1 for
$A_0=2m_{1/2}$ and $A_0=-2m_{1/2}$. We find
that the BR can be observed in the upcoming PSI experiment. In our calculation, 
we satisfy the dark matter relic density constraint i.e.
$0.07<\Omega_{\chi^0_1}h^2<0.25$. The BR of  $\tau\rightarrow
\mu\gamma$ is small.
\begin{figure}\vspace{-0cm}
 \centerline{ \DESepsf(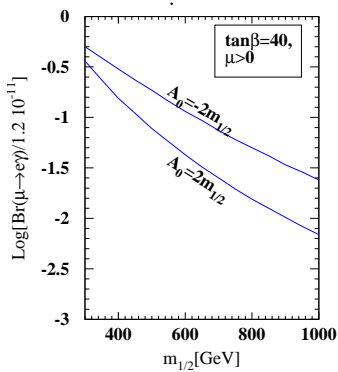 width 3.8 cm) } 
\caption {\label{fig1}$Log_{10}[{Br[\mu\rightarrow e\gamma]\over{1.2\times
10^{-11]}}}]$ as a function of $m_{1/2}$}
\end{figure}
This work was supported in part by the National Science
Foundation Grant PHY - 0101015.
 
\end{document}